\documentclass[%
 reprint,
superscriptaddress,
 amsmath,amssymb,
 aps, pre,
]{revtex4-2}
\usepackage{bbm}
\usepackage{physics}
\usepackage{subcaption}
\usepackage{graphicx}
\usepackage{dcolumn}
\usepackage{bm}
\usepackage[mathlines]{lineno}
\usepackage{float}
\usepackage{xcolor}
\usepackage{physics}
\usepackage{listings}
\usepackage{physics}
\usepackage{amsmath}
\usepackage{array}
\usepackage{printlen}
\usepackage{multirow}

\usepackage[utf8]{inputenc}
\usepackage[english]{babel}
\usepackage[normalem]{ulem}
\usepackage{wasysym}

\usepackage{caption}
\newcommand{\hrho}{\hat{\rho}}

\newcommand{\BF}{\textbf}

\newcommand{\MC}{\mathcal}

\newcommand{\EQi}{\begin{equation}}
\newcommand{\EQf}{\end{equation}}

\begin{document}
\preprint{APS/123-QED}
\title{Dynamical Susceptibilities Near Ideal Glass Transitions}

\author{Corentin C.L. Laudicina}
\affiliation{Soft Matter \& Biological Physics, Department of Applied Physics, Eindhoven University of Technology,
P.O. Box 513, 5600MB Eindhoven, The Netherlands
}%
\author{Chengjie Luo}
\affiliation{Soft Matter \& Biological Physics, Department of Applied Physics, Eindhoven University of Technology,
P.O. Box 513, 5600MB Eindhoven, The Netherlands
}%
\author{Kunimasa Miyazaki}
\affiliation{Department of Physics, Nagoya University, Nagoya 464-8602, Japan}%
\author{Liesbeth M.C. Janssen}%
\affiliation{Soft Matter \& Biological Physics, Department of Applied Physics, Eindhoven University of Technology,
P.O. Box 513, 5600MB Eindhoven, The Netherlands
}%

\email{contact: l.m.c.janssen@tue.nl}

\date{\today}

\begin{abstract}
Building on the recently derived inhomogeneous mode-coupling theory, we extend the generalised mode-coupling theory of supercooled liquids to inhomogeneous environments. This provides a first-principles-based,  systematic and rigorous way of deriving high-point dynamical susceptibilities from variations of the many-body dynamic structure factors with respect to their conjugate field. 
This new framework allows for a novel and fully microscopic possibility to probe for collective relaxation mechanisms in supercooled liquids near the mode-coupling glass transition.
\end{abstract}

\maketitle

\section{Introduction}
Despite decades of intense research efforts, the nature of the glass transition remains a largely open problem in solid state physics. One of the most intriguing facets of glass forming liquids is the emergence of dynamical heterogeneities at the onset of glass formation \cite{kob1997dynamical,donati1998stringlike,marcus1999experimental,kegel2000direct,russell2000direct,glotzer2000spatially,doliwa2000cooperativity,weeks2000three,lavcevic2003spatially,berthier2011dynamical_book,zhang2011cooperative,kim2013multiple,wisitsorasak2014dynamical,tah2021understanding}. These dynamic structures arise as correlated clusters of transiently mobile particles coexisting with regions of immobile particles. The existence of these correlated clusters is in line with the earliest phenomenological approaches to the glass transition of Adam-Gibbs \cite{adam1965temperature}, where the concept of `cooperatively rearranging regions' was first introduced. Dynamical heterogeneity appears to be a universal feature amongst glass formers \cite{berthier2011dynamical_book}, and has also been observed e.g.\ in athermal systems near the jamming point \cite{dauchot2005dynamical,avila2014strong} and in dense active systems \cite{park2015unjamming,malinverno2017endocytic,janssen2019active, janzen2021aging}. More recently, studies have focused on the morphology of these correlated clusters, determining preferential structural ordering in fast and slow moving clusters \cite{boattini2020autonomously}, as well as their fractal dimension and compactedness \cite{appignanesi2006democratic,starr2013relationship,smessaert2013distribution}. Yet, the understanding of dynamical heterogeneity, its origin and consequences with regards to the observed experimental dynamical arrest stand as important missing pieces in the physics of the glass transition.

The established standard measure of dynamical heterogeneity takes the form of a four-point dynamical susceptibility denoted $\chi_4(t)$ \cite{kob1997dynamical,glotzer2000spatially,doliwa2000cooperativity,lavcevic2003spatially,berthier2011dynamical_book,zhang2011cooperative,kim2013multiple,wisitsorasak2014dynamical,tah2021understanding}. In an analogy with critical phenomena, $\chi_4(t)$ is generally defined as the variance of some mobility field, say $\mu(\BF{r}, t)$ which quantifies the mobility of a particle initially located at position $\BF{r}$ and time $t$. Dynamical heterogeneity is captured by the covariance of such a mobility field in both space and time. Mathematically we may write $\operatorname{cov}[\mu(\BF{r}, t)\mu(\BF{r}', t')] \propto G_4(\BF{r}, t ; \BF{r}', t') = G_4(\BF{r}-\BF{r}', t-t')$ using spatial and time translation invariance. The four-point susceptibility is then be given by the spatial integration of the latter:

	\begin{equation}
	\begin{split}
	\chi_4(t-t') &= \int d\BF{r} G_4(\BF{r}, t-t')  \\ &\sim \int d\BF{r} B(\BF{r},t-t') e^{-\abs{\BF{r}}/\xi_4(t-t')},
	\end{split}
	\label{X4_scaling}
	\end{equation}
where $\xi_4(t)$ is the dominant dynamical length-scale associated with the heterogeneities and $B(\BF{r},t-t')$ some scaling function \cite{berthier2011dynamic}. The name four-point stems from the fact that the mobility field is generally a two-point function already \cite{berthier2011dynamical_book}. Numerous studies (both experimental and numerical) \cite{marcus1999experimental,kegel2000direct,lacevic2003approach, vogel2004temperature,berthier2011dynamical_book} demonstrate that such four-point susceptibilities grow in a peak-like manner as the experimental glass transition is approached. The peak is generally identified to occur for times of order $t\sim\mathcal{O}(\tau_{\alpha})$ with $\tau_{\alpha}$ representing the structural relaxation time scale of the system. The fact that dynamical heterogeneity manifests itself in both space and time implies that on time scales $t \gg \tau_{\alpha}$ the four-point susceptibility decays to zero. However, persistence of dynamical clusters beyond $\tau_{\alpha}$ have also recently been reported \cite{kim2013multiple}.

In the mildly supercooled regime, a wide array of glass-forming materials display a power law growth  $\chi_4(\tau_{\alpha}) \sim \tau_{\alpha}^{1/\gamma}$ with a smooth crossover to a logarithmic growth $\chi_4(\tau_{\alpha}) \sim \ln(\tau_{\alpha})^{1/\psi}$ close to the experimental glass transition, see \cite{dalle2007spatial} and references therein. Unfortunately, obtaining an accurate measure of $\chi_4(t)$ in the supercooled regime is a complicated task both computationally and experimentally. Indeed, measuring four-point susceptibilities requires long temporal (several decades) and extensive spatial resolutions (sub-particle radii resolution over large length-scales) since one is probing for correlations in spontaneous fluctuations of the mobility field. Furthermore, some have reported that this four-point susceptibility is highly anisotropic \cite{flenner2009anisotropic}, while most studies consider angularly averaged susceptibilities. The four-point susceptibilities have also been found to bear non-trivial dependences on the choice of both dynamics and statistical ensemble in which simulations are performed \cite{berthier2007spontaneous, berthier2007spontaneousb}. It would be of interest to have a more consistent way to measure dynamical heterogeneity spanning all approaches to the glass problem: theories, simulations and experiments.

\section{Theory}

Equation \eqref{X4_scaling} defines the dynamic length scale $\xi_4(t)$ over which the dynamics are strongly correlated in time. While the existence of a strictly diverging dynamic length scale at the structural glass transition remains debated in the field, it is clear that there exists at least an emerging and growing one from the combination of numerical and experimental results available \cite{weeks2002properties,keys2007measurement,weeks2007short,flenner2011analysis,flenner2014universal}. This provides a basis for a viable critical phenomena-like description of glass formation \cite{andreanov2009mode} and is the motivation for the present work. 

Within the Landau formalism of second order phase transitions, susceptibilities of order parameters are defined as a functional variation of the latter with respect to their conjugate field. A common order parameter to detect the glassy phase is the collective intermediate scattering function $F_2(\BF{k}, t) \propto \langle \hrho_{\BF{k}}^*(0)\hrho_{\BF{k}}(t)\rangle_0$ where $\hrho_{\BF{k}}(t)$ is a collective density fluctuation mode at wavevector $\mathbf{k}$ and time $t$:

    \begin{equation}
        \hrho_{\BF{k}}(t) = \sum_{j=1}^N e^{i\BF{k}\cdot\BF{r}_j(t)} - \left\langle \sum_{j=1}^N e^{i\BF{k}\cdot\BF{r}_j(t)} \right\rangle_0.
    \end{equation}
The statistical averaging $\langle ... \rangle_0$ is performed with respect to the Hamiltonian of a homogeneous (i.e.\ field-free and thus translationally invariant) system. By interpreting the equation of motion for this order parameter as a dynamical Landau theory, the associated susceptibility is obtained by considering variations of this order parameter with respect to its conjugate field that we denote $U(\BF{q}_0)$. The external field $U(\BF{q}_0)$ can be thought of as a small pinning field, or equivalently as a source term in the partition function. This dynamic susceptibility, denoted as $\vartheta_3(t)$, reads 

	\begin{equation}
	\lim_{U\rightarrow0} \frac{\delta F_2(\BF{k} ; \BF{k}', t)}{\delta U(\BF{q}_0)} = \vartheta_3(\BF{k} ; \BF{k}', \BF{q}_0, t)
	\label{chi3_def}
	\end{equation}
where $F_2(\BF{k} ; \BF{k}', t) \propto \langle \hrho_{\BF{k}}^* \hrho_{\BF{k}'}(t)\rangle$ is written with two explicit wave number modes, because the presence of the pinning field $U$ breaks translational invariance. This type of three point susceptibility was first proposed as an alternative to $\chi_4(t)$ \cite{biroli2004diverging,bouchaud2005nonlinear,biroli2006inhomogeneous, berthier2007spontaneous, berthier2007spontaneousb}. The four-point and three point susceptibilities are intimately linked quantities. If the conjugate field is assumed to be the density field, a simple combination of the Cauchy-Schwartz inequality and the fluctuation dissipation theorem leads to the conclusion that $\chi_4(t) \geq \rho_0 \kappa_T \beta^{-1} (\vartheta_3(t))^2$ \cite{berthier2005direct,berthier2007spontaneous}, where $\rho_0$ is the bulk-fluid density, $\kappa_T$ the isothermal compressibility and $\beta$ the inverse temperature of the system under consideration. Similar expressions can be derived in the case of different conjugate fields. 

The three point susceptibility has the advantage of being much easier to determine than the four-point one, as obtaining good enough statistics to take the derivative of the intermediate scattering function is easier than obtaining a statistically relevant four-point correlation function. Following its introduction, several groups were able to extract this three point susceptibility from high-precision dielectric spectroscopy experiments in canonical glass forming materials \cite{berthier2005direct,crauste2010evidence, brun2011nonlinear, bauer2013cooperativity, casalini2015dynamic,mishra2019dynamic}, as well as in molecular dynamics simulations \cite{brambilla2009probing,kim2013dynamic}. In this light, the mode-coupling theory (MCT) \cite{bosse1978mode, gotze2008complex,janssen2018mode} of the glass transition was then extended to the presence of external fields \cite{biroli2006inhomogeneous} and, interpreting it as a Landau theory \cite{andreanov2009mode}, an equation of motion for the three point susceptibility Eq.~\eqref{chi3_def} was derived.

Here we consider multi-point susceptibilities that arise when going beyond the standard MCT. For this we use the framework of so-called generalised mode-coupling theory (GMCT) \cite{szamel2003colloidal, janssen2015microscopic}, which is an extended theory that has been shown to quantitatively improve on some of the weaknesses of MCT \cite{szamel2003colloidal,wu2005high, janssen2015microscopic, luo2020generalised, luo2020generalised2,debets2021generalised,ciarella2021multi,luo2021tagged}. These weaknesses can, in part,  be traced back to an \textit{uncontrolled approximation} made to obtain a self-consistent equation of motion for $F_2(\BF{k},t)$. The GMCT effectively delays this uncontrolled approximation by including physical higher-order density correlations in the picture. Such higher-order density correlation functions are defined as straight generalisations to $F_2(\BF{k},t)$: $F_{2n}(\BF{k}_1, ... , \BF{k}_{n}, t) \propto \langle \hrho_{\BF{k}_1}^*...\hrho_{\BF{k}_n}^*\hrho_{\BF{k}_1}(t)...\hrho_{\BF{k}_n}(t)\rangle$. This property of the GMCT which enables the treatment of an arbitrary number of multi-point correlation functions $F_{2n}$, provides direct access to the higher-order dynamical susceptibilities defined as straight generalisations to Eq.~\eqref{chi3_def}:

    \begin{equation}
    \begin{split}
\vartheta_{2n+1}&(\BF{k}_1,...,\BF{k}_n ; \BF{k}_1,...,\BF{k}_n, \BF{q}_0, t)\\& \equiv \lim_{U\rightarrow0} \frac{\delta F_{2n}(\BF{k}_1,...,\BF{k}_n ; \BF{k}_1,...,\BF{k}_n, t)}{\delta U(\BF{q}_0)}.
    \end{split}
    \end{equation}
These new collective responses are expected to shed light on heterogeneous dynamics near the GMCT transition line. In particular, these susceptibilities provide a way to systematically study isolated responses of many-body relaxation processes to controlled perturbations in both space and time, something which has been missing from the conventional treatments of dynamical heterogeneity.

The paper is organised as follows: in the next part, the GMCT is extended to inhomogeneous environments by appropriately considering a spatially varying external field $U(\mathbf{r})$ chosen to couple to the density modes. This is a completely novel result, and this new framework is referred to as inhomogeneous generalised mode-coupling theory (IGMCT). By treating the equations of motion in reciprocal space for the $2n$-point correlators as a Landau theory \cite{andreanov2009mode}, their functional variations with respect to the external field $U(\BF{q}_0)$ then results in an equation of motion for the odd-point susceptibilities $\vartheta_{2n+1}(t)$ of any order which can in turn be solved self-consistently once a closure relation for IGMCT has been determined. Then, the developed equations of motion are mapped onto a simplified self-consistent relaxation model inspired by earlier studies of schematic mode-coupling theories. A detailed numerical study of the simplified models is also provided.

\subsection{Inhomogeneous Generalised Mode-Coupling Theory}
Consider a classical Newtonian interacting fluid composed of $N_p$ particles in the presence of an arbitrary spatially varying external field $U(\BF{r})$ that modulates the density field $\rho(\mathbf{r})$. Let $\langle ... \rangle$ denote statistical averaging with respect to the inhomogeneous Hamiltonian. Working in Fourier (reciprocal) space, the first set of dynamical variables of interest for GMCT are time-dependent density fluctuation multiplets, denoted as $A^{(n)}_{\mathbf{k}_1,...,\mathbf{k}_n}(t) \equiv \prod_{j=1}^n\hat{\rho}_{\mathbf{k}_j}(t)$ where $n=1,2,...$ and $\hat{\rho}_{\mathbf{k}}(t) = \sum_{l=1}^{N_p} e^{i\mathbf{k}\cdot\mathbf{r}_l(t)} - \langle \rho_{\mathbf{k}} \rangle$ is a density fluctuation. The second set of dynamical variables of interest are the projected density currents, proportional to the time derivative of the fluctuation multiplet:  $\dot{A}^{(n)}_{\mathbf{k}_1,...,\mathbf{k}_n}(t)$. By construction, at equal times (taken to be zero without loss of generality), the correlation of density-density multiplets is proportional to the $2n$-body static structure factor $S_{2n}$:  $\langle A^{(n)}_{\mathbf{k}_1,...,\mathbf{k}_n}(0)^*A^{(n)}_{\mathbf{k}_1',...,\mathbf{k}_n'}(0)\rangle = N_pS_{2n}(\mathbf{k}_1,...,\mathbf{k}_n ; \mathbf{k}_1',...,\mathbf{k}_n')\delta\left(\sum_{j=1}^n(\mathbf{k}_j-\mathbf{k}_j')-\mathbf{q}_0\right)$, while the proper autocorrelation of density-density multiplets is proportional to the $2n$-body dynamic structure factor denoted $F_{2n}$: $\langle A^{(n)}_{\mathbf{k}_1,...,\mathbf{k}_n}(0)^*A^{(n)}_{\mathbf{k}_1',...,\mathbf{k}_n'}(t)\rangle = N_pF_{2n}(\mathbf{k}_1,...,\mathbf{k}_n ; \mathbf{k}_1',...,\mathbf{k}_n', t)\delta\left(\sum_{j=1}^n(\mathbf{k}_j-\mathbf{k}_j')-\mathbf{q}_0\right)$. Here $\mathbf{q}_0$ is the wave vector corresponding to the localised perturbation induced by the external field $U(\mathbf{q}_0)$, and the Dirac $\delta$-function is included to enforce total momentum conservation. For reasons of notational clarity, these momentum enforcing $\delta$-functions will be omitted in the results below. Furthermore a condensed notation for function arguments is used: $\{\BF{k}_j\}_{l:m/p} \equiv (\BF{k}_l, \BF{k}_{l+1}, ..., \BF{k}_{m-1},\BF{k}_m)$ in which $\BF{k}_p$ is omitted.

Using the standard Mori-Zwanzig projector formalism \cite{zwanzig2001nonequilibrium, reichman2005mode,grabert2006projection, gotze2008complex} [see Supplementary Information (SI)], the following equation of motion for the $2n$-body dynamic structure factors can be derived: 
    
    \begin{widetext}
    \begin{equation}
    \begin{split}
0 &= \ddot{F}_{2n}(\{\BF{k}_j\}_{1:n} ; \{\BF{k}_{j}\}_{n+1:2n}, t) + \nu_{2n}\dot{F}_{2n}(\{\BF{k}_j\}_{1:n} ; \{\BF{k}_j\}_{n+1:2n}, t) + \sum_{i=1}^n\int d\BF{k} \Omega^2(\BF{k}_i ; \BF{k}) F_{2n}(\BF{k}_i, \{\BF{k}_j\}_{1:n / i} ; \{\BF{k}_j\}_{n+1:2n}, t) \\
       &+ \frac{1}{n!}\int_0^td\tau\int d\{\BF{k}_j'\}_{1:n} \MC{M}_{2n}(\{\BF{k}_j\}_{1:n} ; \{\BF{k}_j'\}_{1:n}, t-\tau)\dot{F}_{2n}(\{\BF{k}_j'\}_{1:n} ; \{\BF{k}_j\}_{n+1:2n}, \tau),
    \end{split}
    \label{iGMCT}
    \end{equation}
    \end{widetext}
where the quantity $\Omega^2(\BF{k} ; \BF{k}')$ is commonly referred to as a bare frequency (due to its physical dimension). The bare frequency term can explicitly be written as
    \begin{equation}
    \begin{split}
	\Omega^2(\BF{k} ; \BF{k}') = \frac{k_BT}{m}\int d\BF{p} (\BF{k}\cdot\BF{p})\phi(\BF{k}-\BF{p})S^{-1}(\BF{p} ; \BF{k}'),
	\end{split}
    \label{frequency_iGMCT}
    \end{equation}
where $\phi(\BF{k}) = N_p^{-1}\langle \hrho_{\BF{k}} \rangle$ is the average of a single density mode. Equation \eqref{frequency_iGMCT} above has a form very similar to that of homogeneous GMCT \cite{janssen2015microscopic} and reduces to it upon taking the zero-field limit. Indeed, in both the homogeneous and inhomogeneous cases, the bare frequencies of higher-order correlators are simply equal to a sum of two-body bare frequencies. 

The integral kernel $\mathcal{M}_{2n}(t)$ is known as the $2n$-th memory function. To first order it is possible to show that $\mathcal{M}_{2n}(t) \propto F_{2(n+1)}(t)$ \cite{szamel2003colloidal, janssen2015microscopic}. This is precisely what is meant by a hierarchy of coupled integro-differential equations that successively involves higher-order density correlation functions. Explicitly, the memory kernel reads: 
    
    \begin{widetext}
    \begin{equation}
    \begin{split}
        &\mathcal{M}_{2n}(\{\BF{k}_j\}_{1:n} ; \{\BF{k}_j''\}_{1:n}, t)\\
        &= \frac{1}{((n+1)!)^2n!}\int d\{\BF{k}_j'\}_{1:n}d\{\BF{q}_j\}_{1:n+1}d\{\BF{q}_j'\}_{1:n+1} \MC{V}_{2n}^{\dagger}(\{\BF{k}_j\}_{1:n} ; \{\BF{q}_j\}_{1:n+1})F_{2(n+1)}(\{\BF{q}_j\}_{1:n+1} ; \{\BF{q}'_j\}_{1:n+1}, t) \\
        &\hspace{2.5cm}\times \MC{V}_{2n}(\{\BF{k}_j'\}_{1:n} ; \{\BF{q}_j'\}_{1:n+1})J_{2n}^{-1}(\{\BF{k}_j'\}_{1:n} ; \{\BF{k}_j''\}_{1:n}),
    \end{split}
    \label{memory_iGMCT}
    \end{equation}
    \end{widetext}
where the time independent factors $\mathcal{V}_{2n}$ are c-numbers commonly refered to as vertices. These factors represent the effective coupling strengths between different density modes, hence the name 'mode-coupling'.
The detailed form of the vertices is presented in the SI.

In the vertex terms as well as the bare frequency term calculations, a Gaussian factorisation \cite{janssen2015microscopic} of $2n$-body static structure factors and the corresponding inverses is used: $S_{2n}^{(-1)}(\{\mathbf{k}_j\}_{1:n} ; \{\mathbf{k}_j'\}_{1:n}) \approx S_2^{(-1)}(\mathbf{k}_1 ; \mathbf{k}_1')...S_2^{(-1)}(\mathbf{k}_n ; \mathbf{k}_n') + (n-1)!$  permutations over $\{\mathbf{k}_j'\}_{1:n}$. The vertices also contain another type of $n$-body static density correlations which are expressed as $S_{2n+1}(\{\mathbf{k}_j\}_{1:n} ; \{\mathbf{k}_j'\}_{1:n+1}) = N_p^{-1}\langle A^{(n)}_{\mathbf{k}_1,...,\mathbf{k}_n}(0)^* A^{(n+1)}_{\mathbf{k}_1',...,\mathbf{k}_{n+1}'}(0) \rangle$. In the homogeneous GMCT \cite{szamel2003colloidal,janssen2015microscopic}, such correlation functions are factorised in an ad-hoc fashion using a mixture of three-body convolution \cite{barrat1988equilibrium,bosse1978mode} and Gaussian factorisation approximations \cite{janssen2015microscopic}. They are not factorised here to keep the equations of motion as formally exact as possible in light of the linear response expansion that follows. The last term of Eq.~\eqref{memory_iGMCT}, $J_{2n}^{-1}$, is the inverse equal time correlation of the density currents.

Expression \eqref{iGMCT} then forms a hierarchy of coupled equations of motion for many-body dynamic structure factors for a glass forming monatomic liquid in the presence of an external field. While the formal limit $n\rightarrow\infty$ can be considered, it is common to self-consistently close the hierarchy at finite order by approximating $\mathcal{M}_{N_c}(t)$ at some hierarchy height $N_c$ using products of lower-order many-body dynamic structure factors. This is what is called a `mean-field closure' in the generalised mode-coupling literature \cite{janssen2015microscopic, luo2020generalised, luo2020generalised2}. By setting $n=1$ and applying the mode-coupling approximation $F_4(\BF{k}_1,\BF{k}_2 ; \BF{k}_1',\BF{k}_2', t) \approx F_2(\BF{k}_1 ; \BF{k}_1', t)F_2(\BF{k}_2 ; \BF{k}_2', t) + (\BF{k}_1' \leftrightarrow \BF{k}_2')$, the inhomogeneous MCT originally derived in \cite{biroli2006inhomogeneous} is appropriately recovered. For the detailed derivation of IGMCT at arbitrary order, we refer the reader to the SI. 

Interestingly, taking the zero field limit ($U(\mathbf{q}_0) \rightarrow 0$) of Eq.~\eqref{iGMCT} gives direct access to so-called off-diagonal GMCT (oGMCT) \cite{ciarella2021relaxation}, which is a  first-principles improvement upon the original GMCT. 

Indeed, the homogeneous GMCT derived in \cite{janssen2015microscopic} only contains symmetric, angularly averaged $n$-body dynamic structure factors of the form $F_{2n}^{(0)}(k_1, ..., k_n ; k_1, ..., k_n, t)$, thus neglecting the many-body correlations that correspond to physical processes where inelastic scattering of excitations would take place. oGMCT is an attempt to remedy this approximation by including off-diagonal modes in the theoretical framework. This flavor of GMCT was first derived self-consistently for $F_2^{(0)}$ and $F_4^{(0)}$ in \cite{ciarella2021relaxation}. A detailed study of oGMCT would provide important insights on structural glass formers and even possibly remedy the remaining deficiencies of \cite{wu2005high,janssen2015microscopic,luo2020generalised}, but the numerical evaluation of the full oGMCT framework is currently computationally infeasible.
    
\subsection{Dynamical Susceptibilities}

In order to obtain the equations of motion for the many-body response functions $\vartheta_{2n+1}(\mathbf{q}_0;t)$ one considers the variation of Eq.~\eqref{iGMCT} with respect to the external field. We recall that the response of the many-body dynamic correlation function is defined as $ \vartheta_{2n+1}(\mathbf{q}_0; t) \equiv \lim_{U\rightarrow0} \delta / \delta U(\mathbf{q}_0) F_{2n}(t)$. The conjugate variable to the field $U$ is assumed to be a density mode $\hat{\rho}_{\mathbf{q}_0}$ \cite{miyazaki_unpub,kim2013dynamic}. By considering the functional variation of Eq.~\eqref{iGMCT} with respect to the external field, it is easy to see that we obtain a linear hierarchy of coupled equations for the dynamical susceptibilities $\vartheta_{2n+1}(\mathbf{k}_1,...,\mathbf{k}_n ;\mathbf{k}_1',...,\mathbf{k}_n', \mathbf{q}_0,t)$ which inherits the structure of the GMCT hierarchy. Indeed, the $(2n+1)$-th response depends on the $(2(n+1)+1)$-th one since the variation of the memory kernel $\MC{M}_{2n}$ will result in $\MC{M}^{(\vartheta)}_{2n}(t) \propto \vartheta_{2(n+1)+1}(t)$. The self-consistency of the equations of motion for the response functions is uniquely determined by the chosen self-consistent closure of the GMCT hierarchy Eq.~\eqref{iGMCT}. In full, the $(2n+1)-th$ susceptibility reads 

    \begin{widetext}
    \begin{equation}
    \begin{split}
        &\ddot{\vartheta}_{2n+1}(\{\BF{k}_j\}_{1:n} ; \{\BF{k}_j\}_{n+1:2n}, \BF{q}_0,t) + \nu_{2n}\dot{\vartheta}_{2n+1}(\{\BF{k}_j\}_{1:n} ; \{\BF{k}_j\}_{n+1:2n}, \BF{q}_0,t) + \sum_{i=1}^n \frac{|\BF{k}_i|^2}{S(k_i)}\vartheta_{2n+1}(\BF{k}_i,\{\BF{k}_j\}_{1:n / i} ; \{\BF{k}_j\}_{n+1:2n}, \BF{q}_0,t)\\
        &+ \frac{1}{n!}\int_0^t d\tau \int d\{\BF{k}_j'\}_{1:n} \MC{M}^{(\vartheta)}_{2n}(\{\BF{k}_j\}_{1:n} ; \{\BF{k}_j'\}_{1:n}, \BF{q}_0, t-\tau)\dot{F}_{2n}(\{\BF{k}_j'\}_{1:n} ; \{\BF{k}_j\}_{n+1:2n}) \\
        &+ \frac{1}{n!}\int_0^t d\tau \int d\{\BF{k}_j'\}_{1:n}\MC{M}_{2n}(\{\BF{k}_j\}_{1:n} ; \{\BF{k}_j'\}_{1:n}, t-\tau)\dot{\vartheta}_{2n+1}(\{\BF{k}_j'\}_{1:n} ; \{\BF{k}_j\}_{n+1:2n}, \BF{q}_0, \tau) = \MC{T}_{2n}(\{\BF{k}_j\}_{1:n} ; \{\BF{k}_j\}_{n+1:2n}, \BF{q}_0, t),
    \end{split}
    \label{dynamical_susceptibilities_EOM}
    \end{equation}
    \end{widetext}
where the general functional $\mathcal{T}_{2n}$ on the right hand side consists of all terms generated by the field variation which do not contain any susceptibilities (see SI). The equation of motion for the dynamical susceptibility is subject to the momentum conservation constraint $\sum_{j=1}^n \BF{k}_j - \sum_{j=n+1}^{2n} \BF{k}_j - \BF{q}_0 = 0$. 

The left hand side of Eq.~\eqref{dynamical_susceptibilities_EOM} has the form of a linear integro-differential operator, and may be compactly rewritten as $\hat{\mathcal{D}}_{2n}*\vartheta_{2n+1} = \mathcal{T}_{2n}$ where the term containing $\MC{M}_{2n}^{(\vartheta)}$ has been temporarily absorbed in $\mathcal{T}_{2n}$. We expect that the inversion of this operator $\hat{\mathcal{D}}_{2n}$ can lead to divergent behaviour at the critical point, as is the case for conventional critical phenomena. Work along this line was previously done for the closely related four-point susceptibility in a field theoretic setting \cite{bouchaud2005nonlinear}, whose operator actually corresponds to $\hat{\mathcal{D}}_2$ in the present case \cite{bouchaud2005nonlinear,biroli2006inhomogeneous, miyazaki_unpub}. Two of the terms contained in $\MC{T}_{2n}$ are linked to the response of the vertices (see SI). In the IMCT case, it was shown that $\delta \MC{V}_2/ \delta U \sim 0$ \cite{biroli2006inhomogeneous, miyazaki_unpub}. Given that the form of the vertices $\MC{V}_{2n}$ amounts to linear combinations of $\MC{V}_2$ in a homogeneous setting \cite{janssen2015microscopic}, this result may be generalised within a suitable set of ad-hoc factorisation ansatze, such that the variation of vertices of any order is null: $\delta \MC{V}_{2n}/ \delta U \sim 0$. This implies that the very same `mode-coupling' mechanisms are responsible for both the behaviour of the many-body correlators and their associated susceptibilities. A detailed analysis of the operator $\hat{\MC{D}}_{2n}$ and particularily its inversion is however beyond the scope of this work.

The microscopic three-point susceptibility with truncation at hierarchy height $N_c=1$ has been numerically studied in the past \cite{szamel2009three,szamel2010diverging}. It was found that the dynamical length scale associated with $\vartheta_3(t)$ diverges as the ideal MCT transition is approached, confirming initial results \cite{biroli2006inhomogeneous} regarding the scaling behaviour of the IMCT equations. In addition, similar three-point susceptibilities have also been measured in molecular dynamics simulations \cite{brambilla2009probing,kim2013dynamic} as well as in dense colloidal systems \cite{mishra2019dynamic}, all of which provide results that are qualitatively consistent with IMCT predictions. We also mention that the fifth-order susceptibilities recently measured in standard glass formers \cite{albert2016fifth} are distinct from the ones derived in this work as the latter originate from variations of many-body correlators with respect to a conjugate field whereas the former are non-linear susceptibilities originating from an expansion of the polarisibility tensor (whence the notation $\vartheta_{2n+1}$, instead of $\chi_{2n+1}$). 

\section{Insights from a Zero-Dimensional Limit}

\subsection{Setting up the Simplified Model}

\begin{figure*}
    \centering
    \includegraphics[width=\textwidth]{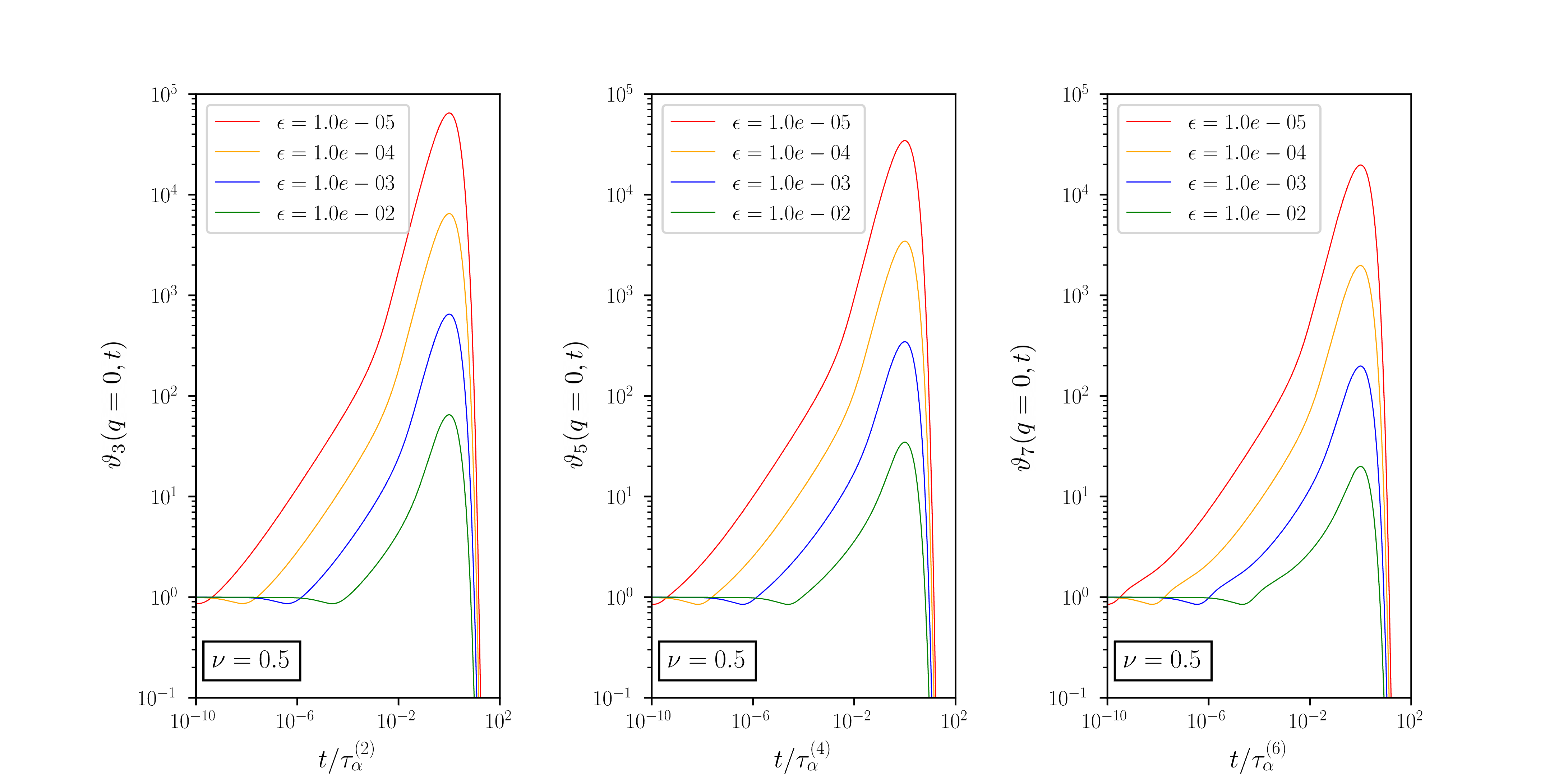}
    \caption{Multi-point dynamical susceptibilities $\vartheta_3$, $\vartheta_5$, $\vartheta_7$ at different distances $\epsilon$ to the critical point from the liquid regime with coupling parameter $\nu=0.5$ and closure [$1^{N_c}$] for hierarchy heights $N_c = 1,\ 2,\ 3$ in left, center, and right panels, respectively. The time axis is rescaled with respect to $\alpha$-relaxation times $\tau_{\alpha}^{(2n)}$.}
    \label{X_trajectories}
\end{figure*}

It is clear that numerically solving for Eq.~\eqref{dynamical_susceptibilities_EOM} is a Herculean task beyond $n=1$ \cite{szamel2009three}. Indeed, Eq.~\eqref{dynamical_susceptibilities_EOM} requires full solutions to oGMCT as initial conditions, which to this day have yet to be obtained \cite{ciarella2021relaxation}. 
In the spirit of early MCT \cite{bengtzelius1984dynamics,leutheusser1984dynamical, gotze2008complex}
as well as early GMCT \cite{mayer2006cooperativity,janssen2014relaxation, janssen2016generalised} studies, we therefore consider a schematic limit of the two coupled hierarchies  Eqs.~\eqref{iGMCT} and \eqref{dynamical_susceptibilities_EOM}. The schematic approach amounts to restricting all physical processes to a single point in momentum space leading to an effective zero-dimensional, self-consistent relaxation process which captures the temporal behaviour of the correlations of interest. 

These models are obtained by setting the structure factor $S(k) = 1 + A \delta(k-k_0)$ where $k_0$ is the wave-number of the first peak of the structure factor. Although simplified, schematic MCT and GMCT models are physically justified by the fact that the main peak of the static structure factor typically dominates over all other wave vectors \cite{bengtzelius1984dynamics, ciarella2019understanding}; additionally, the solutions of the schematic equations are known to share many qualitative features with the fully microscopic theory and, depending on the model parameterisation, schematic and microscopic mode-coupling theories can be regarded as belonging to the same universality class \cite{gotze2008complex}.

In the overdamped limit, this schematic reduction yields the following two coupled hierarchies of integro-differential equations:

    \begin{equation}
    \begin{split}
    \dot{\phi}_{2n}(t) + \mu_{2n} \phi_{2n}+  \Lambda_{2n}\int_0^t d\tau M^{(\phi)}_{2n}(t-\tau)\dot{\phi}_{2n}(\tau) = 0
	\label{schematic_GMCT_hierarchy}
    \end{split}
    \end{equation}
subject to initial conditions $\phi_{2n}(t=0) = 1$, and
    \begin{widetext}
    \begin{equation}
	\begin{split}
	\dot{\vartheta}_{2n+1}(q ; t)&+ \mu_{2n}\vartheta_{2n+1}(q ; t) + \tilde{\mu}_{2n}\phi_{2n}(t) +\Lambda_{2n}\int_0^t d\tau M_{2n}^{(\phi)}(t-\tau)\dot{\vartheta}_{2n+1}(q ; t)(\tau)\\
	&+\tilde{\Lambda}_{2n}(q)\int_0^t d\tau M_{2n}^{(\vartheta)}(q ; t-\tau)\dot{\phi}_{2n}(\tau)+\Lambda_{2n}\int_0^t d\tau M^{(\phi)}_{2n}(t-\tau)\dot{\phi}_{2n}(\tau) = 0
	\end{split}
	\label{schematic_dyn_susc_hierarchy}
	\end{equation}
	\end{widetext}
subject to initial conditions $\vartheta_{2n+1}(q, t=0) = 1$. To provide a link to the microscopic hierarchies : the `mode-coupling' vertices $\MC{V}$ are effectively mapped to a single coupling parameter $\Lambda_{2n}$ which can be interpreted as being proportional to an inverse temperature. It the case of MCT it can be shown that $\Lambda_2 = S(k_0)k_0A^2 / 8\pi^2\rho_0$ \cite{bengtzelius1984dynamics}. The bare frequencies $\Omega^2$ and their variations are mapped to parameter $\mu_{2n}$ and $\tilde{\mu}_{2n}$ respectively. The functional $\mathcal{T}_{2n}$ is mapped to a schematic integral term (i.e. the last term of \eqref{schematic_dyn_susc_hierarchy}). In the equations of motion for the dynamical susceptibility, a perturbative expansion of form $\tilde{\Lambda}_{2n}(q) \equiv \Lambda_{2n}(1-\Gamma q^2)$ is used for the coupling constant. Following \cite{biroli2006inhomogeneous,nandi2014critical}, this form is derived from symmetry considerations of the perturbative expansion of the critical mode-coupling eigenvalue, namely rotational invariance. This straight generalisation to higher-order correlation functions is reasonable because the many-body correlators exhibit the same relaxation patterns as the two-body one since they are governed by equations of motion with similar mathematical form within the generalised mode coupling theory. We thus expect them to behave similarly near an eventual critical point. The wave number dependence $q$ associated with the infinitesimal perturbation is retained to study the behaviour of the dynamical responses with respect to the `length scale' over which it is applied.

In the remainder of this work, we consider schematic IGMCT models with the same parametrisations as introduced in Ref.\ \cite{janssen2014relaxation} for homogeneous GMCT. We define a glassy state as an ergodicity broken state ; this corresponds to the set of points in parameter space beyond which the correlators $\phi_{2n}(t\rightarrow\infty) \equiv f_{2n} > 0$ no longer decay to zero. We parametrise our hierarchies using a level-dependent coupling such that $\Lambda_{2n} = \Lambda n^{1-\nu}$, with $\Lambda,\nu$ real-valued numbers. Furthermore, we assume that the bare frequencies are given by $\mu_{2n} = n$. The memory kernel of a given correlator is defined as $M_{2n}^{(\phi)}(t) = \phi_{2(n+1)}(t)$. Such parametrisations are known to display qualitatively different behaviours upon variation of $\nu$ \cite{janssen2014relaxation}. Furthermore, in the limit of infinite hierarchy height no ergodicity broken states are found at finite coupling strength \cite{janssen2016generalised}. 

In order to obtain numerical results, we are generally forced to close the hierarchy at a given (arbitrary) height. The simplest closure is to set $\phi_{N_c}(t) = 0$, which is referred to as an exponential closure. This type of closure leads to avoided glass transitions as the correlators always decay to zero \cite{janssen2016generalised}. The other common closure is referred to as a `mean-field' one, and consists in setting $\phi_{N_c}(t)$ equal to a monomial of lower order correlators. More precisely, for the hierarchy height $N_c$ at which self-consistency of the hierarchy is desired, we enforce that $M_{2N_c}^{(\phi)} = \prod_{j=1}^p \phi_{2i_j}(t)$, subject to the constraint $\sum_{j=1}^p i_j = N_c$. In this case, we systematically find that the system exhibits a glass transition at finite coupling strength \cite{janssen2016generalised}. Furthermore, one can numerically show that the correlators $\phi_{2n}$ display distinct asymptotic scaling behaviour near the glass transition for a given mean-field closure \cite{janssen2016generalised}. In the early $\beta$-relaxation regime the correlators $\phi_{2n}(t)$ are governed by a power law decay $\phi_{2n}(t)-f^c_{2n} \propto t^{-a_n}$, while in the late $\beta$ to early $\alpha$ regime one finds an additional power law behaviour decay of the form $f^c_{2n}-\phi_{2n}(t) \propto t^{-b_n}$. Additionally the two relaxation times follow power laws, i.e., $\tau^{(2n)}_{\beta}\sim \epsilon ^{-1/2a_n}$ and $\tau^{(2n)}_{\alpha}\sim \epsilon ^{-\gamma_n}$ with $\gamma_n = 1/2a_n + 1/2b_n$. The exponents $a_n,b_n$ are related by the well-known relation \cite{leutheusser1984dynamical, gotze2008complex, janssen2015microscopic, janssen2016generalised}:

    \begin{equation}
        \frac{\Gamma(1-a_n)^2}{\Gamma(1-2a_n)} = \frac{\Gamma(1+b_n)^2}{\Gamma(1+2b_n)}.
    \end{equation}
In the parametrisations considered here, at any order it can be shown that $a_n = 0.395$ and that $b_n = 1$ \cite{mayer2006cooperativity, janssen2014relaxation, janssen2016generalised}.

The coupled hierarchies \eqref{schematic_GMCT_hierarchy}-\eqref{schematic_dyn_susc_hierarchy} are solved using an adaptation of the time doubling scheme presented in \cite{fuchs1991comments,flenner2005relaxation}. This type of numerical scheme has widely been used to solve mode-coupling like integro-differential equations.

\subsection{Numerical Results}

\begin{figure}
    \centering
    \includegraphics[width=1.1\columnwidth]{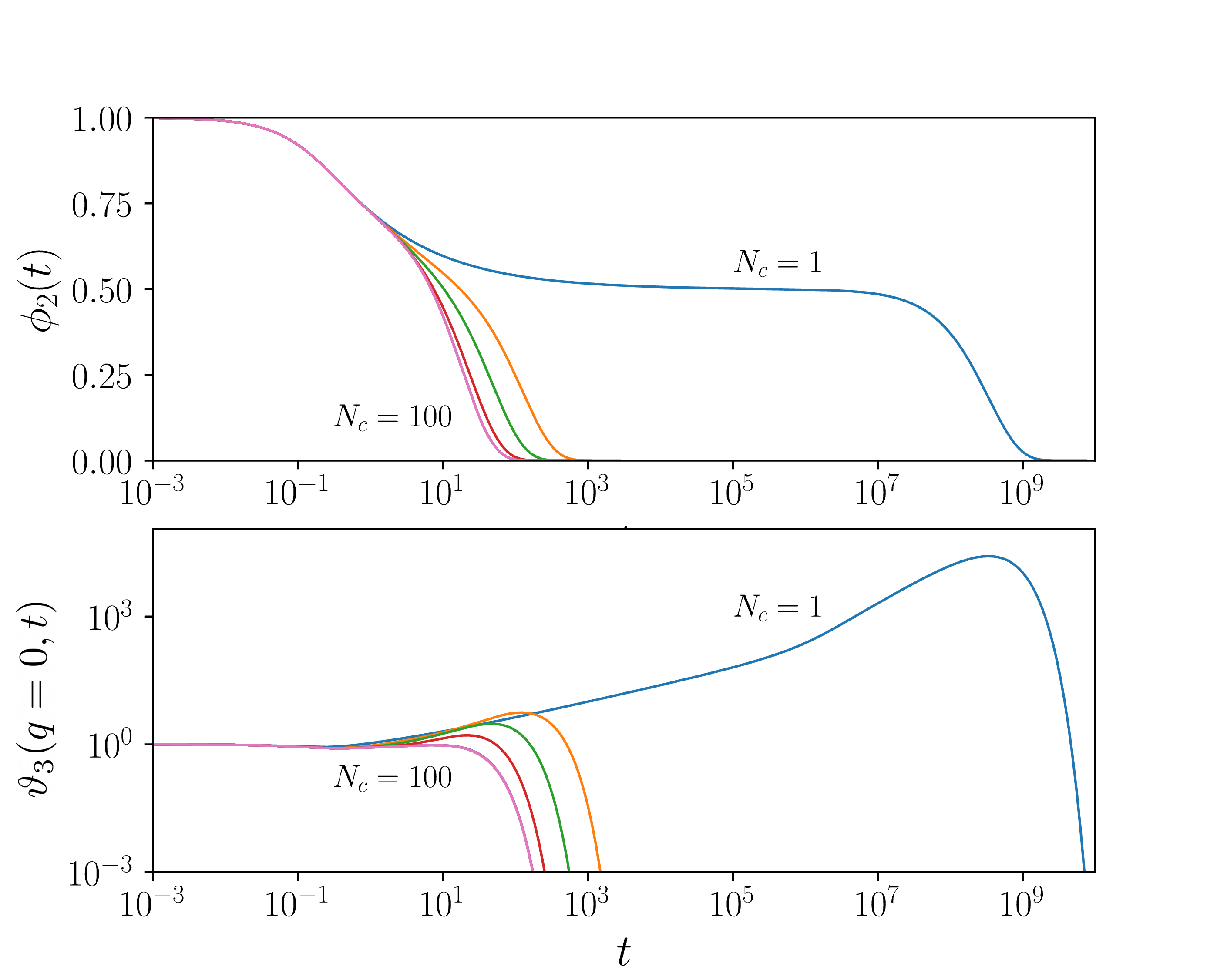}
    \caption{Top panel: principal correlator $\phi_2(t)$ at fixed coupling constant $\Lambda_2 = \Lambda_2^c(N_c=1)-\delta$ with $\delta = 10^{-4}$.  We use a parametrisation with $\nu=1$ and display the correlation function $\phi_2(t)$ for increasing hierarchy height $N_c = 1,2,5,10,100$ appearing in increasing order from uppermost to lowermost curves. Bottom panel: three-point dynamical susceptibility associated with the correlator of the top panel.}
    \label{N_fixed_lbda}
\end{figure}

Similarly to simulation and experimental results \cite{kob1997dynamical,lavcevic2003spatially,berthier2011dynamical_book}, we find that the location of the peak of the nonlinear susceptibilities coincides with that of the $\alpha$-relaxation time $\tau_{\alpha}$ of the corresponding correlation function, as shown in Fig. \ref{X_trajectories} \& \ref{N_fixed_lbda}. For fixed coupling constant $\Lambda_n$, the additional correction levels from the GMCT hierarchy lead to a weakened response to external perturbations as the system is more liquid-like and effectively driven away from criticality by the increasing hierarchy height \cite{mayer2006cooperativity,janssen2016generalised}, as shown in Fig.~\ref{N_fixed_lbda}. This illustrates the principal success of the generalised mode-coupling theory.

The limit $q=0$ corresponding to `macroscopic' perturbations length scales is considered next. In Fig. \ref{X_trajectories}, results for the first three nonlinear susceptibilities $\chi_{3,5,7}(q=0,t)$ obtained from hierarchy heights $N_c = 1,2,3$ respectively, for varying distance to the critical point $\epsilon$ in the liquid regime are shown. There is a marked two-step power law growth $\vartheta_{2n+1}(q=0, t) \propto t^{a_n}$ for $t<\tau_{\beta}^{(2n)}$ while $\vartheta_{2n+1}(q_0, t) \propto t^{b_n}$ for $\tau_{\beta}^{(2n)} < t < \tau_{\alpha}^{(2n)}$ followed by a very fast decay of the dynamical susceptibilities for longer times. These results hold for susceptibilities of any order, and the exponents $a_n$, $b_n$ correspond to the well-known mode-coupling exponents previously mentioned and originally determined in \cite{janssen2015microscopic}.

We sketch a qualitative proof of the above observations. For simplicity we focus on the simpler MCT scenario with a single propagator and a single dynamical susceptibility. Then, $M^{(\phi)}(t) = \phi(t)^2$ and $M^{(\vartheta)}(q ; t) = 2\phi(t)\vartheta(q ; t)$. The argument extends to the GMCT scenario but is less amenable to analytic treatment. Dropping all indices for further clarity, the density propagator in Laplace frequency space reads: $\phi(z) = [z-\mu\Sigma(z)]^{-1}$ with $\Sigma(z) = (1+\Lambda M^{(\phi)}(z))^{-1}$ some (non-perturbative) correction. Similarly the \textit{linear} equation for the dynamical susceptibility can be written as $\vartheta(q ; z)[z+f(q ; z)] = T(z)$ where 
    \begin{equation}
        f(q ; z) = \mu + 2\tilde{\Lambda}(q)\phi(z)\left(z\phi(z)-1\right) + z\Lambda M^{(\phi)}(z),
    \end{equation}
and $T(z)$ some regular function of $z$. It follows that $\vartheta(q ; z)$ grows as $[z+f(q ; z)]$ decreases and eventually diverges once $[z+f(q ; z)] = 0$. It is easy to show that the condition $[z+f(q ; z)] = 0$ is a quadratic equation in $\phi(z)$ whose solution admits the following form $\phi(z) \propto 1/z + \MC{O}(z)$ in the limit $z\ll 1$. Translating back to the time-domain, we have that the dynamical susceptibility $\vartheta(t)$ will grow whenever $\phi(t)$ is slowly varying over macroscopic time scales. This is precisely what happens at the onset of glass-like behaviour where propagators exhibit double step decays with an intermediate plateau, as displayed in the upper panel of Figure \ref{N_fixed_lbda}. This calculation also demonstrates how the dynamical susceptibilities inherit the scaling exponents of the propagators, but with a reversed sign due to the operator inversion. We expect that this argument also extends to the fully microscopic theory, since the mathematical structure of the time dependence is unchanged. \\

\begin{figure}
    \centering
    \includegraphics[width=1.05\columnwidth]{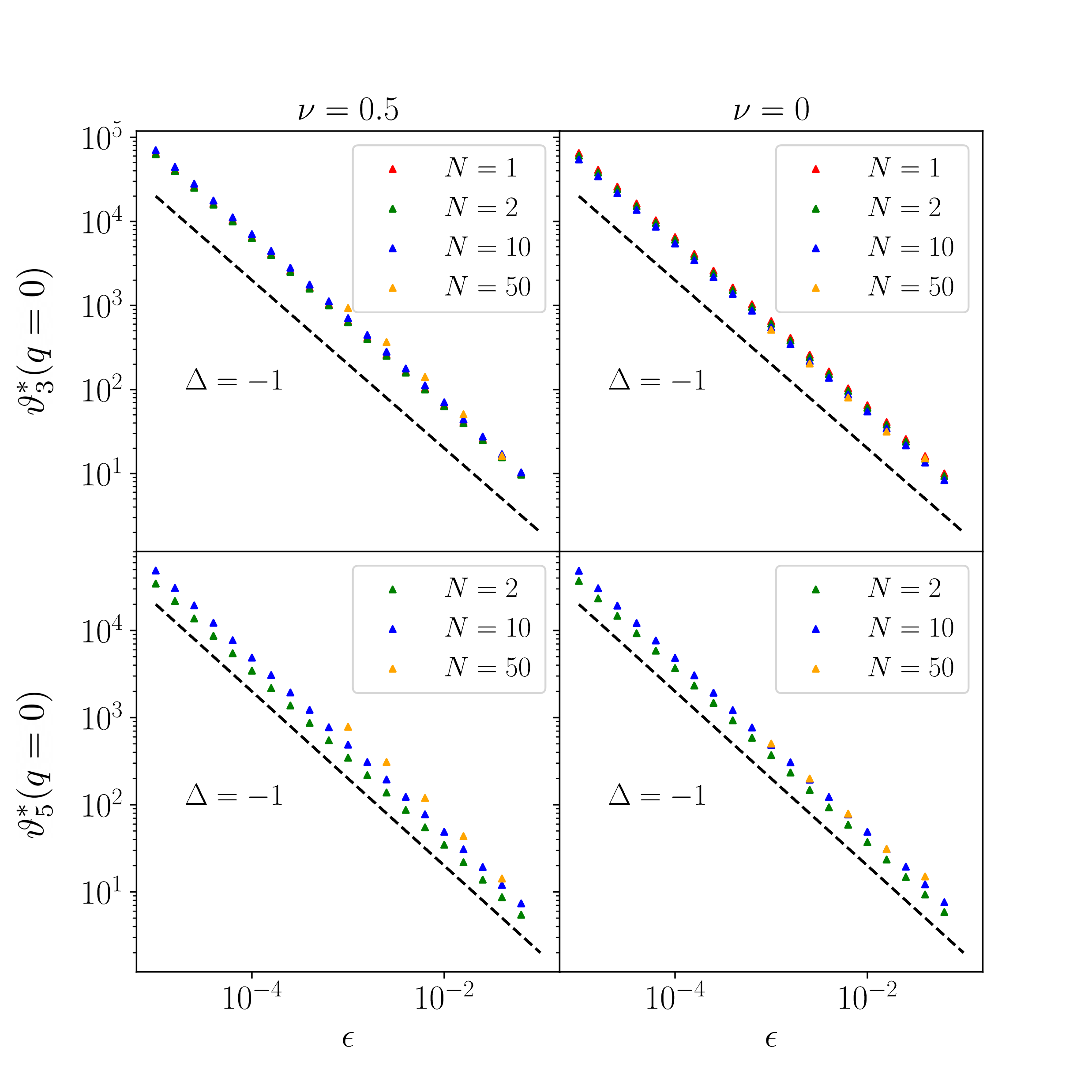}
    \caption{Power law scaling behaviour of the maximum of the first two susceptibilities $\vartheta_3^*(q=0),\vartheta_5^*(q=0)$ as as a function of the distance $\epsilon$ to the glass transition. We display invariance of the exponent with respect to the hierarchy height $N_c = N$ and coupling constant parametrisation $\Lambda_{2n} = \Lambda n^{1-\nu}$. The exponent $\Delta = -1$.}
    \label{scaling_X3_X5}
\end{figure}

Moreover, we find that the peak of the dynamical susceptibilities scales in a similar way to what is found in \cite{dalle2007spatial}. Indeed, the maximum of the dynamical response of any order scales as a power law $\vartheta_{2n+1}^*(q=0) \propto \epsilon^{\Delta}$ with exponent $\Delta = -1$ as the transition is approached, irrespective of the global hierarchy level and of the parametrisation of the coupling coefficients $\Lambda_{2n}$, as displayed in Fig.~\ref{scaling_X3_X5}. Given that the hierarchy Eq.~\eqref{schematic_GMCT_hierarchy} is known to display qualitatively different scaling laws related to fragility \cite{janssen2014relaxation}, this implies that the scaling laws governing the divergence of the nonlinear dynamical susceptibilities are universal near the simplest glass transition, regardless of relaxation behaviour. This universality is expected to hold for all systems exhibiting $\MC{A}_2$-glass singularities, as the equations of motion can be appropriately mapped to their normal form close enough to the transition point \cite{gotze2008complex}. 
Akin to the correlators, the dynamical susceptibilities are found to converge with increasing hierarchy height at fixed coupling. This is in line with prior results on schematic GMCT, which show a manifestly convergent behaviour as the hierarchy height is increased while keeping the coupling strength fixed \cite{mayer2006cooperativity,janssen2014relaxation}. Moreover, numerical results in the case of the homogeneous microscopic theory substantiate this \cite{szamel2003colloidal,wu2005high,luo2020generalised,luo2020generalised2}. We therefore expect that the solutions to the microscopic equations Eq.~\eqref{dynamical_susceptibilities_EOM} eventually converge with increasing hierarchy height.

\begin{figure}
    \centering
    \includegraphics[width=1.05\columnwidth]{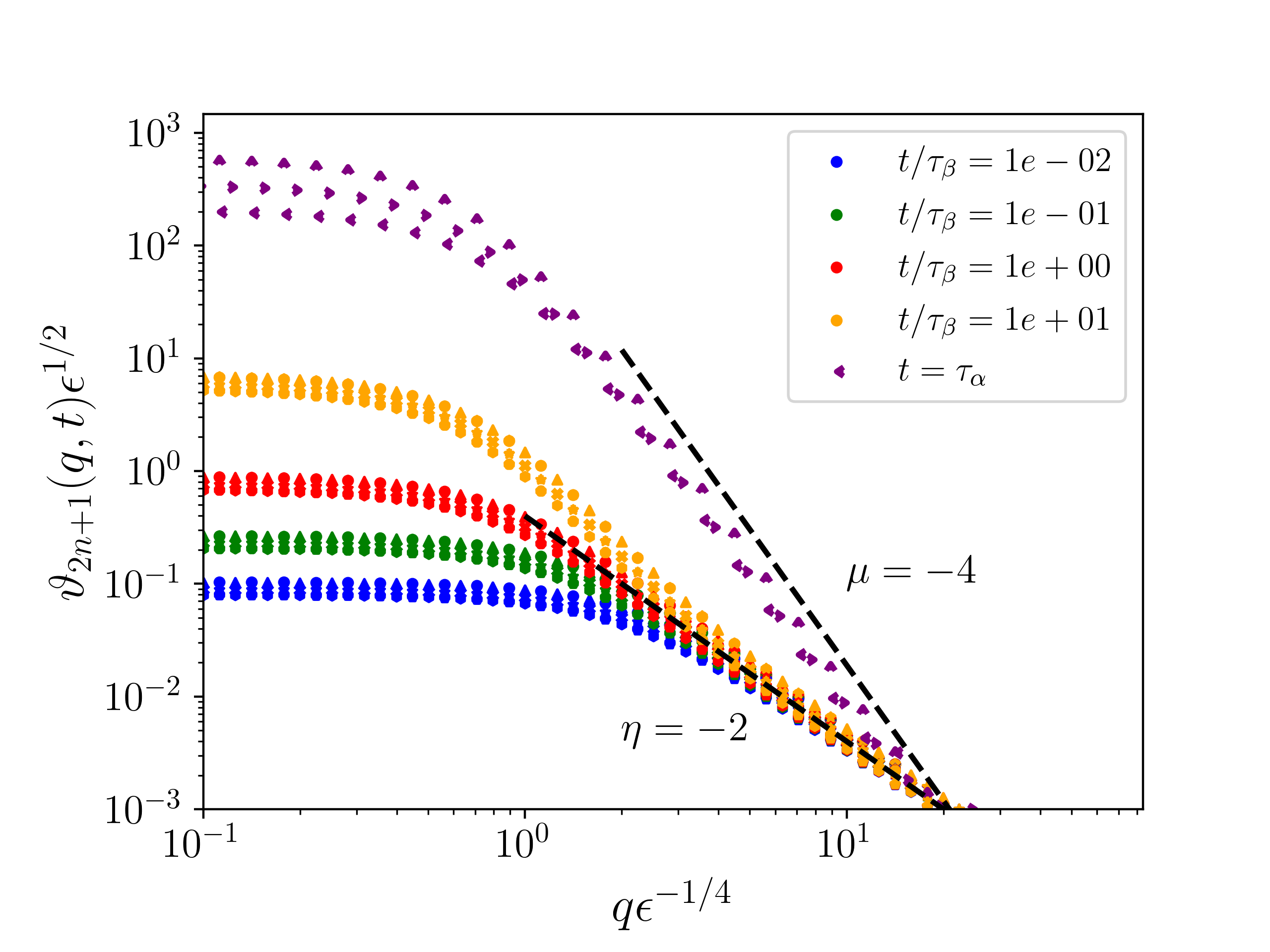}
    \caption{Verification of scaling law Eq.~\eqref{scaling_law} near the $\tau_{\beta}$ regime and at $\tau_{\alpha}$ for the dynamical susceptibilities $\vartheta_3$, $\vartheta_5$ and $\vartheta_7$ obtained for hierarchy heights $N_c=1$, 2, and 3, respectively, with $\nu=0.5$. Symbols: $\vartheta_3 : \{\ocircle, \triangledown, \hexstar \}$, $\vartheta_5 : \{*, \times, + \}$ and $\vartheta_7 : \{\triangleleft, \triangleright, \triangle \}$ correspond to $\epsilon = \{ 10^{-5},\ 10^{-6},\ 10^{-7}\}$. respectively. The slopes $\eta, \mu$ correspond to that of the scaling law Eq.~\eqref{scaling_law}.}
    \label{q_scaling_graph}
\end{figure}

Figure \ref{q_scaling_graph} reports on the magnitude of the dynamical responses at different time scales as a function of the perturbation `length scale' $q^{-1}$. As previously noted, the response grows in time, reaching its peak at the $\alpha$-relaxation time. Furthermore, the magnitude of the response is strongly $q$-dependent at all time scales. Indeed, the dynamical responses reach a maximum for some threshold macroscopic perturbation length scale $q_*$, indicating that these are genuine collective responses. From Fig. \ref{q_scaling_graph}, we see that $q_*$ depends on both the time scales probed as well as the distance to the glass transition. Overall, denoting $\xi_{2n}$ the length scale associated with the $2n+1$-th dynamical susceptibility, a global scaling of the form 

    \begin{equation}
        \vartheta_{2n+1}(q,t) \approx \xi_{2n}^2g_{2n+1}(\xi_{2n}q, t/\tau_{\beta})
    \label{scaling_law}
    \end{equation}
is observed for the nonlinear susceptibilities at all orders $n$. This is a generalisation of the scaling for the higher-order susceptibilities initially determined in \cite{biroli2006inhomogeneous, tarzia2010anomalous, miyazaki_unpub}. The scaling function $g_{2n+1}$ reads 
    
    \begin{equation*}
        g_{2n+1}(x,y) = \frac{1}{\alpha_{n} y^{-a_n} + x^2} + \frac{\beta_{n}y^{b_n}}{1+x^2+\gamma_{n}y^{c_n}x^4},
    \end{equation*}
where $\alpha_n, \beta_n, \gamma_n$ are real-valued coefficients which depend on the choice of parametrisation of the hierarchies Eqs.~\eqref{schematic_GMCT_hierarchy}-\eqref{schematic_dyn_susc_hierarchy}. The prefactor $\xi_{2n}^2$ in \eqref{scaling_law} is a consequence of the Lorentzian form of the first term of the scaling function.  The exponents $a_n, b_n$ are the GMCT exponents previously introduced and $c_n$ governs the growth of the quartic tail seen in Fig.~\ref{q_scaling_graph} beyond the $\beta$-relaxation regime. We numerically determine that $c_n$ lies in the range $(0.15,0.20)$ for the parametrisations considered, in line with earlier results \cite{miyazaki_unpub}. In the limit $x\rightarrow0$ the sequential power-law behaviour $\vartheta_{2n+1}(q=0,t) = \alpha_n^{-1}(t/\tau_{\beta}^{(n)})^{a_n}+\beta_n(t/\tau_{\beta}^{(n)})^{b_n}$ observed in the numerical solutions shown in Fig.~\ref{X_trajectories} and detailed in the text is recovered. In the $\alpha^{(n)}$-relaxation regime (i.e. $t/\tau_{\beta}^{(n)}\gg1$) and at low wave numbers ($q\to0$) one finds, since $a_n = 0.395$ and $b_n=1$, that the peak of the susceptibility must scale according to

    \begin{equation*}
    \begin{split}
        \vartheta_{2n+1}^*(q) &\approx \xi_{2n}^2\beta_n(t/\tau_{\beta}^{(n)})^{b_n} \propto \epsilon^{-\Delta},
        \end{split}
    \end{equation*}
as numerically determined. Using $(t/\tau_{\beta}^{(n)})^{b_n} = \epsilon^{-1/2}$ \cite{janssen2014relaxation, miyazaki_unpub}, we conclude that there exists a set of diverging length scales going as $\xi_{2n} \propto \epsilon^{-1/4}$ at the $\alpha$-relaxation time scale (since $\Delta = -1$). Thus, a set of true diverging dynamical length scales associated with dynamical susceptibilities of any order is expected at the ideal glass transition, where the separation parameter $\epsilon$ vanishes.

Finally, it is interesting to briefly consider the limit $n\rightarrow\infty$ of Eqs.~\eqref{schematic_GMCT_hierarchy}-\eqref{schematic_dyn_susc_hierarchy}. We recall that previous studies demonstrate that this particular limit washes away any critical singularity and that the  $\alpha$-relaxation time scale does not diverge at finite coupling constant for particular parametrisations of the bare frequencies and coupling constants \cite{mayer2006cooperativity, janssen2014relaxation, janssen2016generalised}. This implies that the dynamical susceptibilities introduced here no longer critically diverge at finite coupling constant as the limit of infinite hierarchy height is taken. Nonetheless, analytic treatment of the infinite hierarchy \eqref{schematic_GMCT_hierarchy} have shown that it exhibits the very same MCT power law scalings $a_n, b_n$ and $\gamma_n$ in a limited parameter range \cite{mayer2006cooperativity}. Hence, a quantitative agreement between finite and infinite hierarchies within the same coupling constant window is expected.

\section{Conclusion}

In this work the GMCT of the glass transition has been extended to the presence of an external pinning field that locally modulates the microscopic density. By treating the resulting equations of motion as a Landau theory and making use of the fact that GMCT incorporates an arbitrary number of many-body dynamic structure factors, a set of self-consistent equations of motion for previously unstudied nonlinear dynamical susceptibilities $\vartheta_{2n+1}$ is found. The equations of motion for nonlinear dynamical susceptibilities can be casted into an eigenvalue problem and it is hypothesised that a true diverging susceptibility, and hence a diverging dynamic length scale exists at the (generalised) mode-coupling critical point. This hypothesis is corroborated by earlier results on the fully microscopic mode-coupling theory \cite{bouchaud2005nonlinear,biroli2006inhomogeneous,szamel2009three}, and on analogous simpler self-consistent relaxation models which have shown to capture the phenomenology of the wave-vector dependent MCT equations \cite{leutheusser1984dynamical, gotze2008complex,mayer2006cooperativity,janssen2015microscopic}. Near the simplest glass singularity predicted by mode-coupling theories, diverging susceptibilities of any order asymptotically close to the critical point are found. Akin to \cite{biroli2006inhomogeneous}, the dynamic length scales $\xi_{2n}^*$ of any order at the $\alpha$-relaxation time scale are numerically found to diverge as $\epsilon^{-1/4}$ at the critical point. The value of this exponent is in line with microscopic calculations of MCT \cite{szamel2010diverging} but does not coincide with results from molecular dynamics simulations \cite{flenner2011analysis,kim2013dynamic} which find that the dynamic length scale diverges as $\epsilon^{-1/2}$ in both hard and soft sphere systems. Recent results for fragile liquids however seem to be in line with an MCT-like scenario, where a dynamical length scale diverging as $\epsilon^{-0.3}$ was reported \cite{tah2022kinetic}.
The scaling behaviour of the susceptibilities is studied in both space and time, and in various limits of the physical parameters for the simplified self-consistent relaxation models. The analysis of these models strengthens the idea that these new, higher-order dynamical susceptibilities are capable of capturing dynamical heterogeneity in a previously unexplored manner. The nonlinear dynamical susceptibilities introduced in this work provide us with new tools to make quantitative predictions on this still poorly understood facet of glass physics. We reserve a detailed analytic and numerical study of the scaling laws near the set of higher order glass singularities predicted by the microscopic generalised mode-coupling theory \cite{gotze2002logarithmic, luo2021sticky}, extending on \cite{nandi2014critical}, for future work. Furthermore, it would be interesting to measure these non-linear susceptibilities in molecular glass formers, inhomogeneous MD simulations or in simpler model glass forming systems such as kinetically constrained models as was recently suggested in \cite{biroli2021amorphous}. \\

\textbf{Acknowledgements}: We thank Prof. G. Biroli for showing us the explicit form of the scaling function for the three-point susceptibtility. We thank I. Pihlajamaa for useful comments on the manuscript. The authors acknowledge financial support from the Dutch Research Council (NWO) through a Vidi grant (C.C.L.L. and L.M.C.J.) and START-UP grant (C.L. and L.M.C.J.). \\

\bibliography{bibliography.bib}
\end{document}